%% file: ms.tex
\documentclass[iop]{emulateapj}

\usepackage[colorlinks=true,linkcolor=blue,anchorcolor=blue,citecolor=blue,urlcolor=blue]{hyperref}
\hypersetup{pdfauthor={Jong Chul Lee}, pdftitle={A Submm Survey of Local DOGs}}

\newcommand{\iras}{\textit{IRAS} }

\slugcomment{Last updated: \today}
\shorttitle{A Submm Survey of Local DOGs}
\shortauthors{Lee et al.}

\usepackage{natbib}

\begin{document}

\title{A SUBMILLIMETER CONTINUUM SURVEY OF LOCAL DUST-OBSCURED GALAXIES}

\author{Jong Chul Lee$^{1}$, Ho Seong Hwang$^{2}$, and Gwang-Ho Lee$^{3}$}
\affil{$^{1}$ Korea Astronomy and Space Science Institute, 776 Daedeokdae-ro, 
              Yuseong-gu, Daejeon 34055, Republic of Korea; jclee@kasi.re.kr\\
       $^{2}$ School of Physics, Korea Institute for Advanced Study, 85 Hoegiro,
              Dongdaemun-gu, Seoul 02455, Republic of Korea\\
       $^{3}$ Department of Physics and Astronomy, Seoul National University,
              1 Gwanak-ro, Gwanak-gu, Seoul 151-742, Republic of Korea\\}

\begin{abstract}

We conduct a 350 $\mu$m dust continuum emission survey of 
  17 dust-obscured galaxies (DOGs) at $z$ = 0.05--0.08 
  with the Caltech Submillimeter Observatory (CSO).
We detect 14 DOGs with $S_{350 \mu m}$ = 114--650 mJy and S/N $>$ 3.
By including two additional DOGs with submillimeter data in the literature, 
  we are able to study dust contents for a sample of 16 local DOGs 
  that consists of 12 bump and 4 power-law types. 
We determine their physical parameters with a two-component modified blackbody function model.
The derived dust temperatures are in the range 57--122 K and 22--35 K
  for the warm and cold dust components, respectively.
The total dust mass and the mass fraction of warm dust component are
  3--34$\times10^{7} M_\odot$ and 0.03--2.52\%, respectively.
We compare these results with those of other submillimeter-detected infrared luminous galaxies. 
The bump DOGs, the majority of the DOG sample, 
  show similar distributions of dust temperatures and total dust mass to the comparison sample. 
The power-law DOGs show a hint of smaller dust masses than other samples, 
  but need to be tested with a larger sample. 
These findings support that the reason why DOGs show heavy dust obscuration is not 
  an overall amount of dust content, 
  but probably the spatial distribution of dust therein.

\end{abstract}


\section{INTRODUCTION}

The cosmic star formation rate density peaks at $z\sim2$ and 
  nearly half of the stars in present-day galaxies 
  formed around this epoch (e.g., \citealt{dic03,magn13,beh13}).
Understanding what drives this intense star formation is 
  one of key issues in the study of cosmic star formation history 
  (see \citealt{mad14} for a review).
It is thus important to efficiently identify $z\sim2$ star-forming galaxies 
  to study the physical mechanisms responsible for the intense star formation activity.
Interestingly, many star-forming galaxies in the distant Universe are 
  infrared- and submillimeter-bright
  because a significant fraction of ultraviolet (UV) photons from 
  young massive stars are absorbed by dust and then 
  re-radiated in the infrared and submillimeter wavelengths
  (see \citealt{san96,ken98,cas14}).

Among various methods to select high-redshift dusty star-forming galaxies,
  an optical-to-mid-infrared color cut $R-[24] \geq 14$
  (mag in Vega, corresponding to $S_{24 \mu m}/S_{0.65 \mu m} \geq 982$)
  with a mid-infrared flux density cut $S_{24 \mu m} \geq 0.3$ mJy 
  (\citealt{dey08}; see also \citealt{hou05,fio08,hwa12,rigu15})
  has been widely used because of its simplicity.
The galaxies satisfying this simple criterion are referred to 
  as dust-obscured galaxies (DOGs); 
  they experience heavy dust obscuration as the name suggests 
  (e.g., \citealt{bra07,mel11,pen12})
  and contribute to about 30\% of the total infrared output
  in the Universe at $z\sim2$ (e.g., \citealt{rigu11,cal13}).
Numerical simulations suggest that 
  luminous DOGs result from extremely gas-rich galaxy mergers, 
  while less luminous ones are driven either by mergers or 
  by secular evolution of gas-rich disk galaxies \citep{nar10}.

The DOGs are divided into two categories based on 
  their spectral energy distribution (SED) shape 
  at rest-frame near- and mid-infrared wavelengths:
  ``bump'' and ``power-law'' DOGs \citep{dey08}.
The bump DOGs show a stellar photospheric bump at 1.6 $\mu$m \citep{joh88,far08},
  whereas the power-law DOGs show a monotonically increasing continuum attributed to 
  hot dust component, indicating the presence of active galactic nuclei 
  (AGNs; \citealt{hou05,des09}; but see also \citealt{nar10}).              
In the merger-driven scenario of DOG formation, 
  the DOGs are thought to evolve from bump (i.e., starburst-dominated) to 
  power-law (AGN-dominated) types
  (e.g., \citealt{nar10,bus11}). 
Some very luminous DOGs ($L_{\rm IR}$ $\gtrsim$ $10^{13}$ $L_\odot$) 
  appear as so-called hot DOGs,
  which are mainly powered by deeply buried AGNs 
  (see \citealt{wu12,jon14,ass15,tob16}).
Therefore, studying the DOGs can provide important hints of 
  a possible evolutionary link among high-$z$ galaxies and 
  the connection between star formation and nuclear activity.
However, because of their extreme distances, 
  it is difficult to compare the observational features with model predictions, 
  which is crucial for understanding what makes a DOG have such large dust obscuration.

To study the physical properties of DOGs in detail
  \citet[hereafter HG13]{hwa13a} focused on local analogs of these galaxies that 
  have a wealth of multiwavelength data available
  (see also \citealt{hec05,jun14,gre16,bia16}). 
Using the multiwavelength data 
  from UV to far-infrared that include
  {\it Galaxy Evolution Explorer} ({\it GALEX}; \citealt{mar05}),
  Sloan Digital Sky Survey (SDSS; \citealt{york00}) data release 7 (DR7; \citealt{aba09}),
  {\it Wide-field Infrared Survey Explorer} ({\it WISE}; \citealt{wri10}),
  {\it Infrared Astronomical Satellite} ({\it IRAS}; \citealt{neu84}), and
  {\it AKARI} Space Telescope \citep{mur07},
  HG13 identified 47 DOGs at $0.05<z<0.08$ with extreme flux density ratios
  between mid-infrared ({\it WISE} 12 $\mu$m) and near-UV ({\it GALEX} 0.22 $\mu$m) bands
  (i.e., $S_{12 \mu m}/S_{0.22 \mu m} \geq 892$).
Comparison of local DOGs to other galaxies with lower $S_{12 \mu m}/S_{0.22 \mu m}$
  shows that local DOGs have a relatively large Balmer decrement (H$\alpha$/H$\beta$),
  small optical size, and large elongation.
On the other hand, there are no significant differences in
  specific star formation rate and in large- and small-scale environments between the two sample.

The multiwavelength data for the DOGs in HG13 mainly cover only $\lambda \leq 100$ $\mu$m; 
  there are few data on the `Rayleigh-Jeans' side of the infrared SED peak, 
  which is important to quantify dust properties of galaxies accurately 
  \citep{hwa10b,dal12,sym13}.
Therefore, \citet[hereafter HAG13]{hwa13b} conducted Submillimeter Array (SMA) observations 
  of four local DOGs to probe 880 $\mu$m continuum emissions. 
They derived dust temperatures and masses using a two-component dust model
  (warm and cold dust components associated with
  stellar birth clouds and diffuse interstellar medium, respectively;
  see \citealt{cha00,dun01,sau05,vla05,dac08,wil09})
  and found that the dust properties of local DOGs are similar to 
  those of other infrared luminous galaxies with submillimeter detection.
HAG13 thus concluded that the DOGs are not a distinctive population among dusty galaxies; 
  probably the reason why some galaxies appear as DOGs is not an extremely large dust content, 
  but simply results from a large dust obscuration along the line of sight.

There were only four local DOGs with submillimeter detection in HAG13. 
To test further the idea on the nature of local DOGs 
  using a larger sample with submillimeter detection, 
  we extend the dust continuum emission survey of local DOGs with the 
  10.4 m single-dish antenna of the Caltech Submillimeter Observatory (CSO) in this study.
The structure of this paper is as follows.
We explain the target selection, CSO observations, 
  and data reduction in Section \ref{data}.
We derive physical parameters of dust content 
  in the local DOGs and compare them with those of other submillimeter-detected
  galaxies in Section \ref{results}.
We summarize and discuss the results in Section \ref{discuss}.
Throughout this paper, we adopt the flat $\Lambda$CDM cosmological parameters:
  $H_0 = 70$ km s$^{-1}$ Mpc$^{-1}$, $\Omega_{\Lambda}=0.7$, and $\Omega_{m}=0.3$.

\section{DATA}\label{data}

\subsection{Targets}\label{target}

HG13 identified 47 local DOGs satisfying the criteria of 
  $S_{12 \mu m}/S_{0.22 \mu m} \geq 892$, $S_{12\mu m} >$ 20 mJy, and $0.05 < z < 0.08$.
We first fitted to their multiwavelength data at $\lambda > 6$ $\mu$m with 
  the SED templates and fitting routine of 
  DECOMPIR\footnote{\url{http://sites.google.com/site/decompir}} \citep{mul11}
  to predict 350 $\mu$m flux densities. 
We then selected 26 bright local DOGs 
  with the expected 350 $\mu$m flux densities $\geq$ 100 mJy and 
  with R.A. = 9--18 hours that could be observable in 2014 March and April runs.

\subsection{Observations and Data Reduction}\label{obs}

Among the 26 targets, 17 objects were actually observed with the CSO telescope
  over four nights under moderate weather conditions
  (the opacity at 225 GHz $\tau_{225}$ = 0.04--0.10). 
The total integration time for each object is 20--80 minutes
  by excluding bad scans.
We also observed Ganymede and Callisto for pointing, focusing, and flux calibration 
  at the beginning and middle of each night (i.e., once every three or four hours).
The pointing accuracy is $\sim4\arcsec$ rms and 
  the flux density uncertainty introduced by the calibration error is 
  typically 5--10\%.

The observations were carried out with 
  the second-generation Submillimeter High Angular Resolution Camera
  (SHARC-II; \citealt{dow03})
  that has a bolometer array with $32\times12$ pixels.
The field of view and beam size (FWHM) of SHARC-II at 350 $\mu$m are 
  $2\arcmin.6\times1\arcmin.0$ and $8\arcsec.5$, respectively.
The CSO Dish Surface Optimization System (DSOS; \citealt{leo06}) 
  was activated to minimize the surface imperfections and gravitational deformations.
To map the area around compact sources including our targets and calibrators, 
  the telescope was swept in the standard Lissajous pattern with amplitudes of 
  $\pm20\arcsec$ and $\pm10\arcsec$ in azimuth and elevation, respectively, 
  resulting in a uniform coverage of $115\arcsec\times38\arcsec$.

We reduced the data using the Comprehensive Reduction Utility for SHARC-II 
  (CRUSH) software package\footnote{\url{http://www.submm.caltech.edu/~sharc/crush}}
  \citep{kov06}, version crush-2.20-3. 
The option ``-faint'' in CRUSH was applied
  because targets were faint, but still visible in a single scan.
The output map had a pixel scale of 1$\arcsec$.62 pixel$^{-1}$
  and was smoothed to an effective FWHM of 12$\arcsec$.4
  for optimal detection.
The photometry was conducted with a 20$\arcsec$ diameter aperture,
  large enough to capture the instrumental flux density of each source. 
The sky level and photometric measurement error were calculated
  by computing the mean and rms within $\sim$10 off-source apertures.
We applied the same procedure to the calibrators
  and derived scaling factors that convert the instrumental flux density 
  to a physical flux density.
Among the 17 observed DOGs, 14 ones were detected 
  with signal-to-noise (S/N) $>$ 3 in the synthesis maps.
The other three DOGs were not detected 
  even though we used the option ``-deep'' for very faint sources in the data reduction. 

\begin{figure*}
\center
\includegraphics[width=160mm]{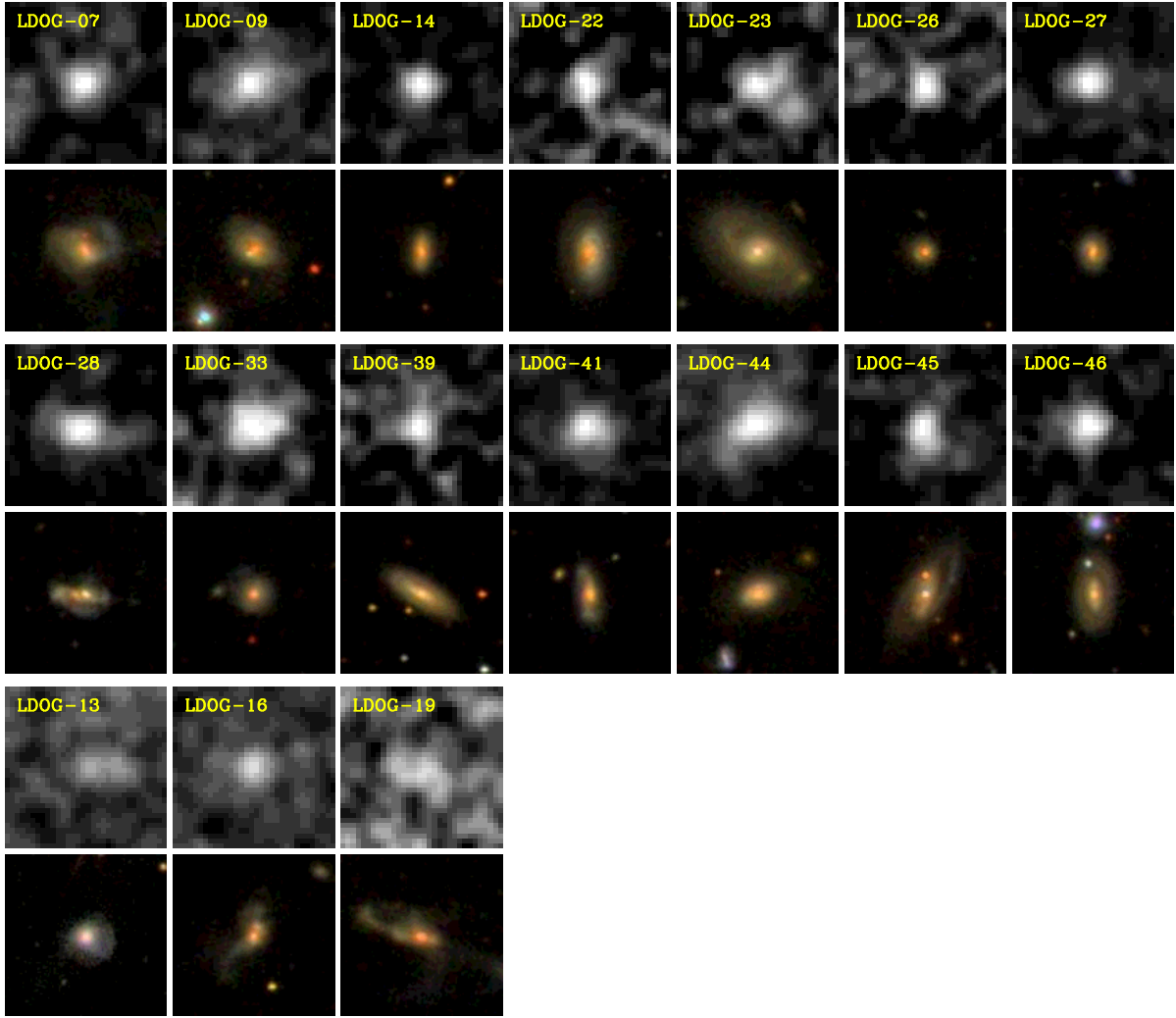}
\caption{
CSO synthesis maps of the 350 $\mu$m continuum emission (1st, 3rd, and 5th rows)
  and SDSS cut-out images (RGB color composites from $irg$ bands; 2nd, 4th, and 6th rows)
  for the 17 CSO-observed DOGs.
The size of each image is $48\arcsec \times 48\arcsec$ (55.6 kpc at $<$$z$$>$=0.06).
North is up and east is to the left.
}\label{fig-cut}
\end{figure*}

We list the CSO observation log of 17 DOGs in Table \ref{tab-samp}
  with their 350 $\mu$m flux densities or 3$\sigma$ upper limits.
The flux density uncertainties include the measurement and calibration errors.
We found the {\it Herschel} 350 $\mu$m photometric data  
  for LDOG$-$07 (609.9$\pm$8.1 mJy) from the {\it Herschel}-ATLAS program \citep{rig11}, 
  which agrees with our measurement (475.9$\pm$86.0 mJy) within 1.6$\sigma$ level.
We display the CSO 350 $\mu$m continuum images in Figure \ref{fig-cut} 
  with the optical color images from SDSS $gri$-band data.
Although some DOGs appear extended in the CSO images, 
  it is not easy to say that they are spatially resolved 
  because the focus correction was not frequently made during observations.

\input{table1}

\section{RESULTS}\label{results}

\subsection{Deriving Dust Temperatures and Masses of Local DOGs}\label{fit}

\begin{figure*}
\center
\includegraphics[width=140mm]{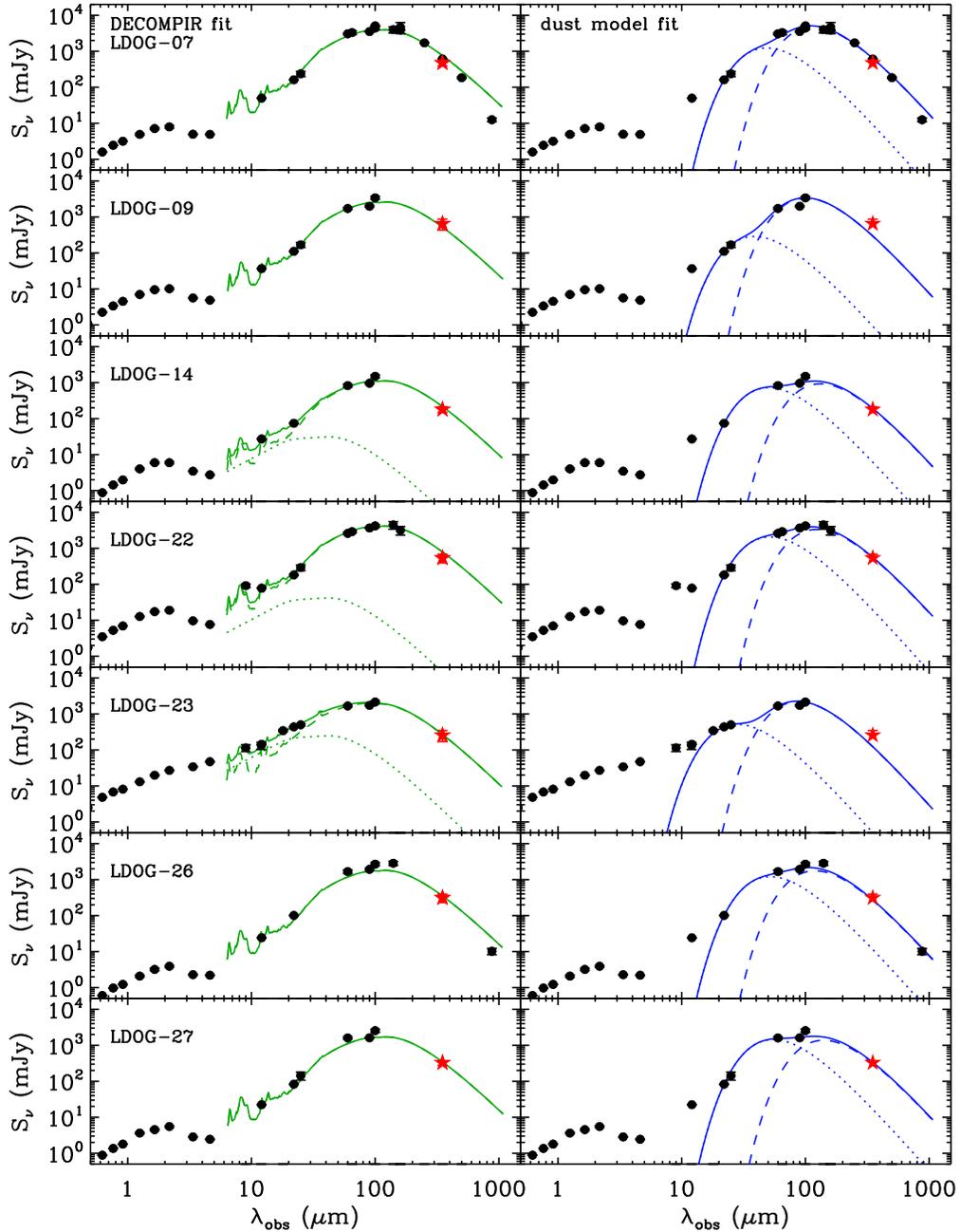}
\caption{
SEDs of the 14 DOGs with CSO detection.
Black circles and downward arrows (for 3$\sigma$ upper limits) are 
  photometric data compiled in HG13 and HAG13,
  while red stars represent our 350 $\mu$m observations.
Error bars are shown for all the points but are mostly smaller than the symbols.
In the left panels, green solid, dotted, and dashed lines indicate the best-fit SEDs
  with the DECOMPIR routine of \citet{mul11}
  for total, AGN and host-galaxy components, respectively.
In the right panels, blue solid, dotted, and dashed lines represent
  the total, warm and cold dust components
  from the two-component modified blackbody function fits.
}\label{fig-sed}
\end{figure*}

\begin{figure*}
\center
\includegraphics[width=140mm]{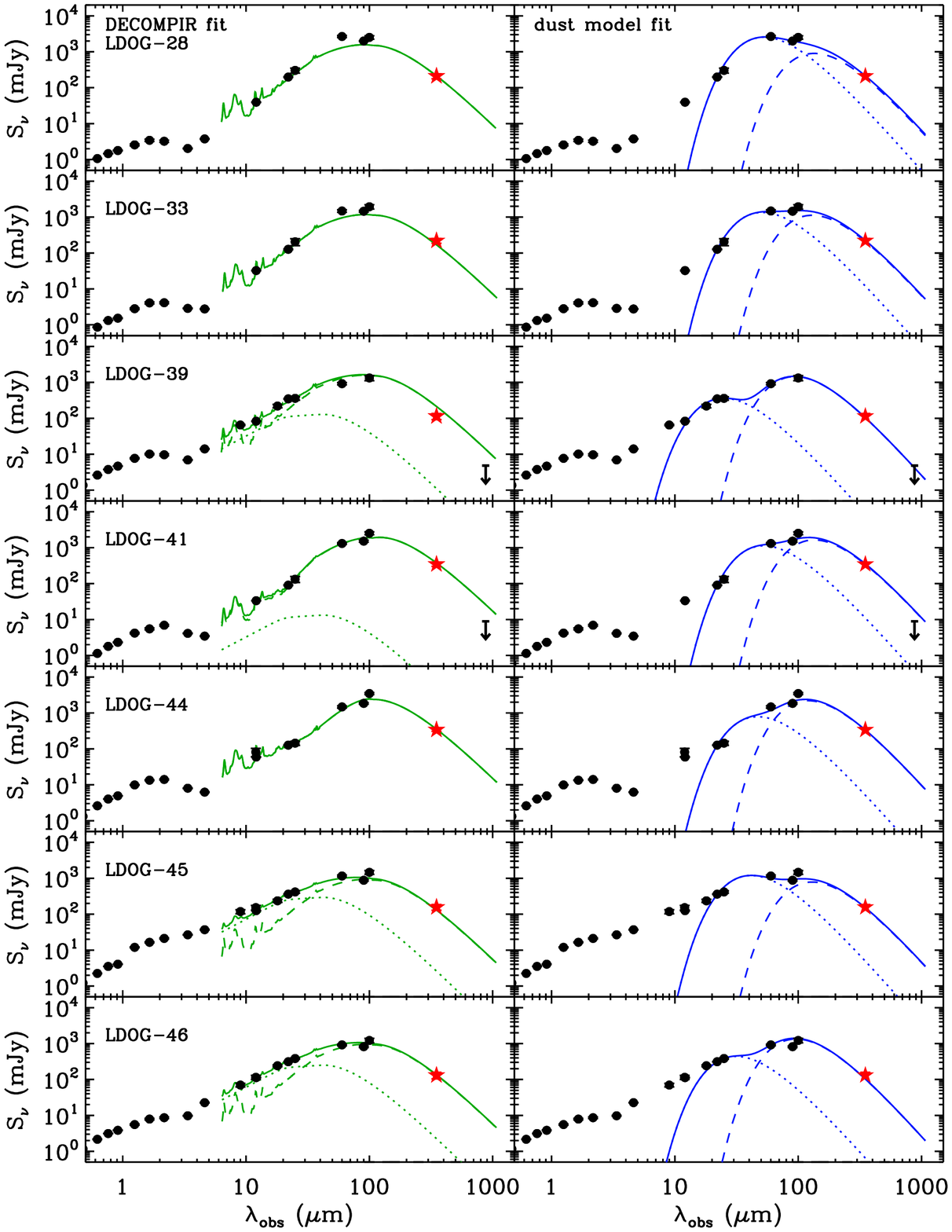}
\begin{center}
Fig. \ref{fig-sed}.--- Continued
\end{center}
\end{figure*}

\input{table2}

We derive the physical parameters of dust content of 
  the 14 local DOGs by fitting to the multiwavelength data 
  including our 350 $\mu$m submillimeter observations.
We again adopt the SED fitting routine of DECOMPIR 
  to derive the total  (8--1000 $\mu$m) infrared luminosities and 
  to decompose the infrared SEDs into AGN and host-galaxy components. 
The DECOMPIR routine contains the SED templates consisting of one AGN SED 
  and five groups of host-galaxy SEDs. 
These templates are produced from the spectroscopic data of 
  {\it Spitzer} infrared spectrograph and 
  the photometric data of {\it IRAS} for AGN-host and starburst galaxies 
  (see \citealt{mul11} for more details). 
We fit to the data at $\lambda > 6$ $\mu$m with these SED templates, and
  choose the best-fit template that provides the lowest $\chi^2$ value for each galaxy.
The left panels of Figure \ref{fig-sed} show the photometric data of 
  the 14 local DOGs with CSO 350 $\mu$m detection and their best-fit SEDs from DECOMPIR. 
We do not use the SMA 880 $\mu$m upper limit flux densities of LDOG$-$39 and LDOG$-$41
  for the SED fits as HAG13 did.
The total infrared luminosities are in the range 
  $1.60\times10^{11} \leq L_{\rm IR}/L_\odot \leq 5.43\times10^{11}$.
The AGN contribution to the infrared luminosity varies 0.0\% to 44.4\%.

We then derive the dust temperatures and masses
  by fitting to the infrared and submillimeter data 
  at $\lambda >$ 20 $\mu$m with a two-component modified blackbody model.
For the optically thin case, this model has the form in the rest-frame:
    \begin{equation}\label{eq-mbb}
    S_{\nu} = A_w \nu^\beta B_\nu(T_{\rm warm}) + A_c \nu^\beta B_\nu(T_{\rm cold}),
    \end{equation} 
  where $T_{\rm warm}$ and $T_{\rm cold}$ are the dust temperatures of 
  warm and cold components, respectively.
  $A_w$ and $A_c$ are the relative contributions
  of the two components, $B_\nu$($T$) is the Planck function, and
  $\beta$ is the dust emissivity index.
The dust emissivity index $\beta$ can vary between 1 and 2 depending on several dust 
  parameters (e.g., dust grain size, composition, temperature; \citealt{dra84,rem13}),
  but it is usually fixed for the fit when the number of data points is small 
  (e.g., $\beta=1.5$: \citealt{bos12,rigu15}; $\beta=2.0$: \citealt{wil09,cor12}).
We test two cases with $\beta=1.5$ and 2.0, 
  and adopt $\beta=2.0$ that provides better fits for our sample.

Then the dust mass can be derived by \citep{hil83}
    \begin{equation}\label{eq-mass}
    M_{\rm dust} = M_{\rm warm} + M_{\rm cold} = \frac{D_L^2~\nu^\beta}{k_\nu}(A_w + A_c),
    \end{equation}
  where $D_L$ is the luminosity distance and 
  $k_\nu$ is the dust mass opacity coefficient 
  (absorption cross section per unit dust mass).
To be consistent with HAG13,
  we adopt $k_\nu$ at 850 $\mu$m ($k_{850}$) = 0.383 cm$^2$ g$^{-1}$ from \citet{dra03}.
This dust mass opacity coefficient $k_\nu$ can be very uncertain, 
  thus the resulting dust mass can change by a factor of two 
  (e.g., $k_{850} = 0.77$ cm$^2$ g$^{-1}$ in \citealt{jam02}).
Comparison of dust masses with other studies should be carefully made 
  by considering different $k_\nu$.
The right panels of Figure \ref{fig-sed} show the best-fit SEDs 
  from the two-component modified blackbody function model. 
Again, we do not use the SMA 880 $\mu$m upper limit flux densities of LDOG$-$39 and LDOG$-$41
  for the SED fits.

In Table \ref{tab-sed}, we list the derived physical parameters including 
  total infrared luminosity, AGN contribution to the infrared luminosity,
  dust temperatures of warm and cold components, total dust mass, 
  and mass fraction of warm dust component.
We determine the uncertainty for each parameter by generating 1000 SEDs
  within the associated photometric error (assumed to be Gaussian distribution)
  and then calculating the standard deviation of SED fitting results.
It should be noted that these uncertainties could be underestimated, 
  especially the total infrared luminosity and the AGN contribution. 
This is because the uncertainties may just reflect the uncertainty in the fitting procedure, 
  not taking fully into account the discrepancy between the data and the best-fit model that 
  could be due to the incompleteness of SED models at longer wavelengths and 
  the difficulty in modeling the intrinsic AGN infrared SEDs.
For the following analysis,
  we add two DOGs that are not observed in this study, 
  but have submillimeter data in the archive: 
  LDOG$-$08 and LDOG$-$35 (see HAG13 for details).

\subsection{Comparison of Dust Properties between
  Local DOGs and Other Infrared Luminous Galaxies}

To compare the dust properties of local DOGs with other infrared luminous galaxies, 
  HAG13 constructed a comparison sample of galaxies with submillimeter detection 
  in the literature. 
This sample contains 62 galaxies with SCUBA 850 $\mu$m data 
  from the SCUBA local universe galaxy survey (SLUGS: \citealt{dun00, dun01}) 
  and/or SMA 880 $\mu$m data from \citet{wil08}.
Their mid- and far-infrared data are adopted from HG13 and 
  from the Great Observatories All-sky LIRG Survey (GOALS: \citealt{arm09, u12}).
The galaxies in the comparison sample do not satisfy the color criterion of DOGs. 
HAG13 also removed very nearby galaxies at $z \leq 0.01$ and 
  interacting systems from the comparison sample (see HAG13 for more details).

There have been several submillimeter data available for local dusty star-forming galaxies 
  since HAG13 (e.g., \citealt{cle13,cie14,cla15}). 
However, we did not add these galaxies in the comparison sample 
  because most of them have $L_{\rm IR} \lesssim 10^{11}$ $L_\odot$ 
  that are different from the infrared luminosity range of local DOGs. 
Although we apply the same SED fitting method to 
  both local DOGs and galaxies in the comparison sample (hereafter comparison galaxies),
  it is not always straightforward to compare them
  because of their inhomogeneous selection criteria.
We therefore focus mainly on relative differences of several dust properties.

\begin{figure}
\center
\includegraphics[width=85mm]{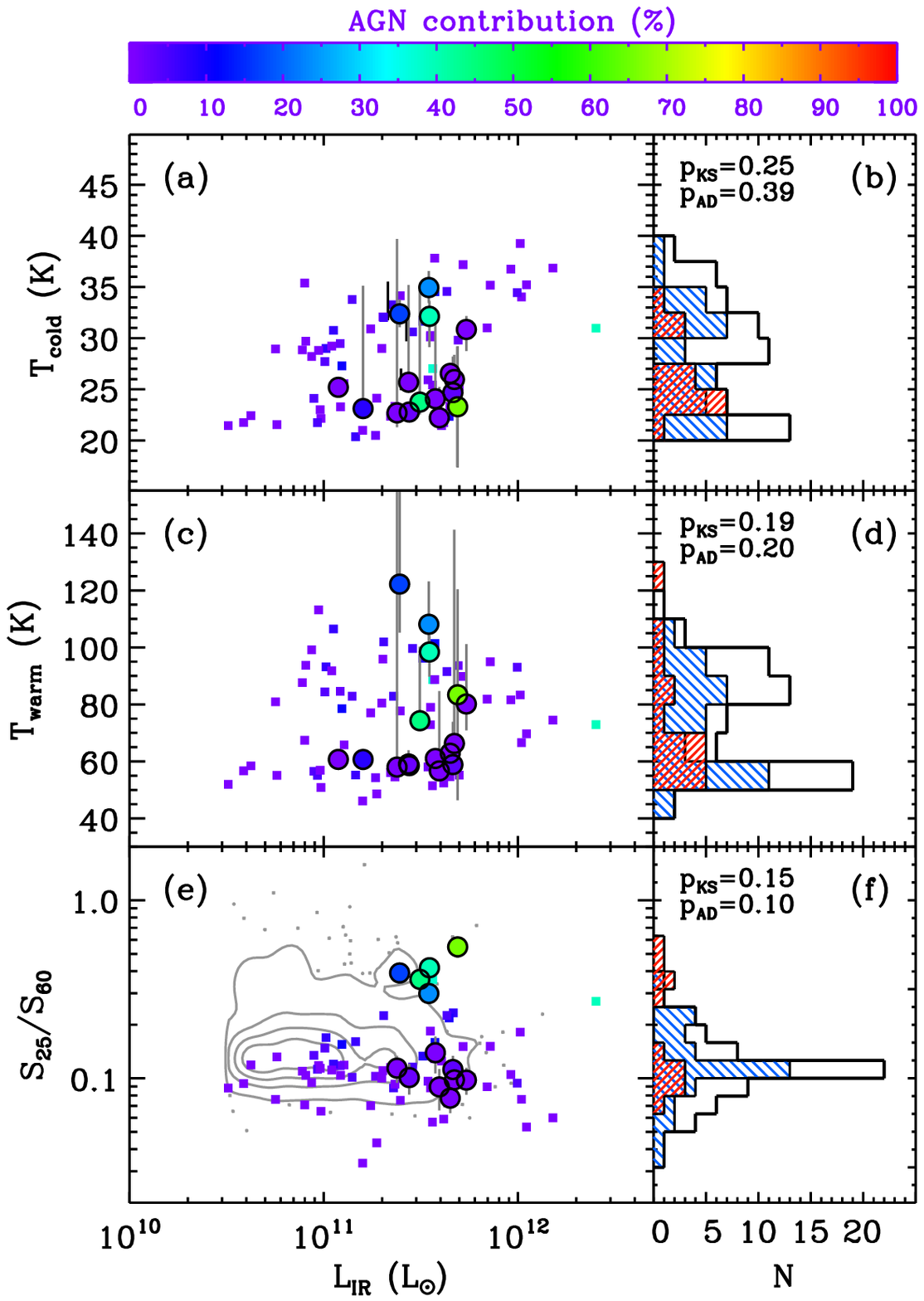}
\caption{
Dust temperature of the cold component
  for local DOGs (circles) and 
  for other infrared luminous galaxies with submillimeter detection (squares)
  as a function of total infrared luminosity (a), and their histograms (b).
Symbols are color-coded as shown by the color bar on the top
  to represent AGN contributions estimated from the SED decomposition.
Error bars are plotted only for local DOGs for better visibility.
The thick solid line histogram represents the distribution of all squares.
The red ($//$) and blue ($\setminus\setminus$) hatched histograms
  denote the local DOGs and the galaxies in the comparison sample 
  in the same range of infrared luminosity 
  $L_{\rm IR}$=1.19--5.43$\times10^{11}L_\odot$, respectively.
Two numbers in the corner are $p$-values from the K-S and A-D k-sample tests 
  between the two distributions.
Same as (a--b), but for the dust temperature of the warm component (c--d) and
  for the flux density ratios between \iras 25 and 60 $\mu$m (e--f).
The gray dots and contours in panel (e) indicate the distribution of 
  {\it IRAS}-detected SDSS galaxies at $z > 0.01$ regardless of submillimeter detection.
}\label{fig-tdust}
\end{figure}

\begin{figure}
\center
\includegraphics[width=85mm]{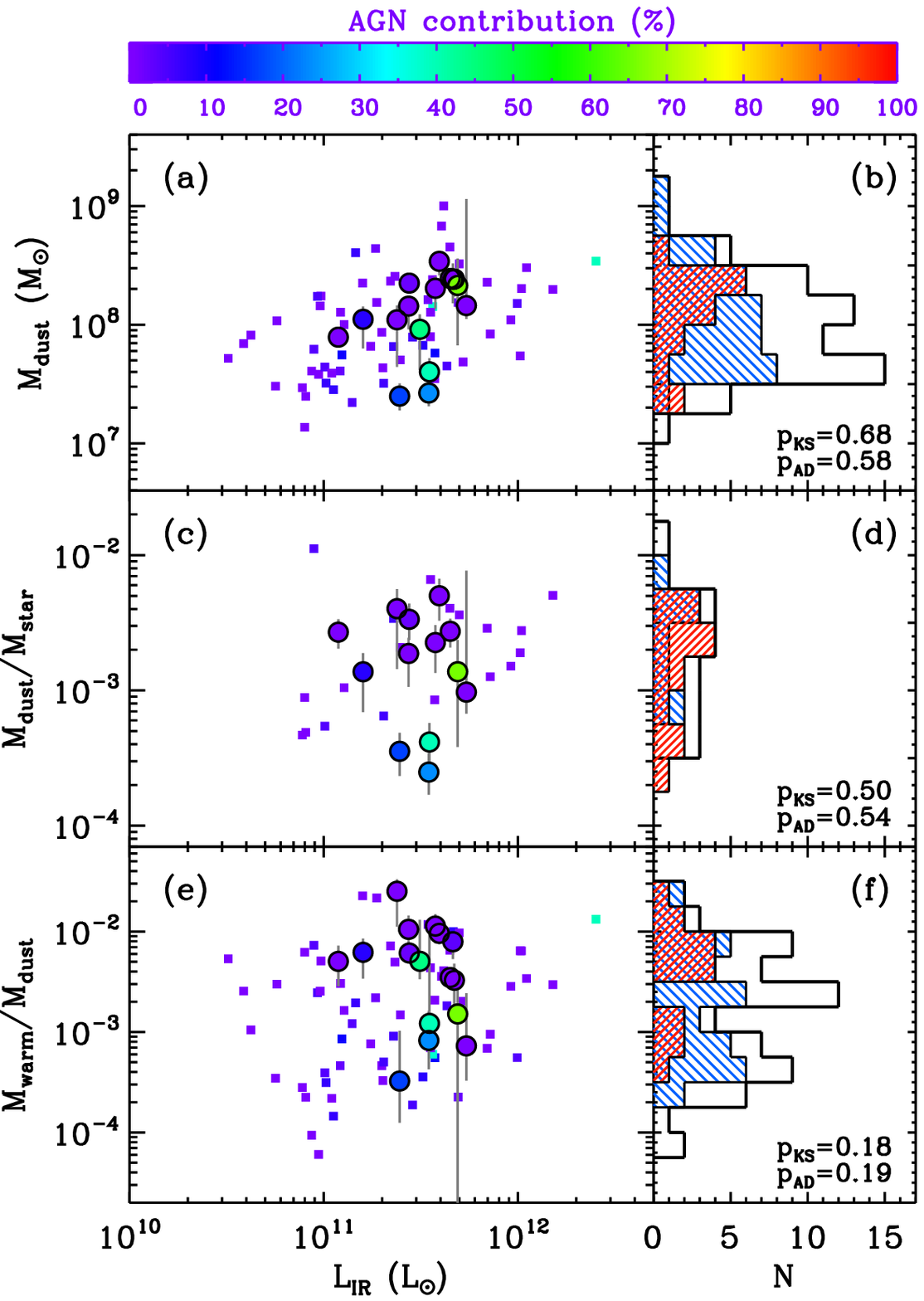}
\caption{Same as Figure \ref{fig-tdust}, but for
the total dust mass (a--b),
the dust-to-stellar mass ratio (c--d), and
the dust mass fraction of warm component (e--f).
}\label{fig-mdust}
\end{figure}

Figures \ref{fig-tdust} and \ref{fig-mdust} display several parameters related 
  to the dust temperature and mass as a function of total infrared luminosity.
The large circles and small squares denote the local DOGs and 
  the comparison galaxies, respectively.
Each symbol is color-coded according to the AGN contribution to infrared luminosity.
Open histogram is for all the comparison galaxies
  regardless of their infrared luminosities.
The red and blue histograms show the distributions for 16 local DOGs 
  and 35 comparison galaxies in the same range of infrared luminosity
  ($1.19\times10^{11} \leq L_{\rm IR}/L_\odot \leq 5.43\times10^{11}$).
For fair comparison, we mainly discuss the difference between the red and blue histograms.
We run the Kolmogrovo-Smirnov (K-S) test and the Anderson-Darling (A-D) k-sample test 
  on the distributions between the two samples, 
  and list the relevant p-values\footnote{
  The $p$-value is involved in the probability that two test samples are extracted
  from the same parent population. In general, a $p$-value $<$ 0.05 is considered
  that the two distributions are significantly different.} in each figure.

The dust temperature and mass of local DOGs appear not to change with infrared luminosity, 
  but the dynamic range for the infrared luminosity is very small ($\sim$0.7 dex).
For the comparison galaxies,
  the cold dust temperature $T_{\rm cold}$ is well correlated 
  with infrared luminosity (Spearman's rank correlation coefficient = 0.47 
  and the probability of obtaining the correlation by chance = 0.03\%),
  while the warm dust temperature $T_{\rm warm}$ is not (see \citealt{dun01}).
The total dust mass $M_{\rm dust}$ shows a good correlation with infrared luminosity
  (Spearman's rank correlation coefficient = 0.50 and 
  the probability of obtaining the correlation by chance $<$ 0.01\%),
  consistent with previous studies (e.g., \citealt{dun01,mag12}).
The dust-to-stellar mass\footnote{
  The stellar masses are drawn form the MPA/JHU DR7 
  value-added galaxy catalogue \citep{kau03}.
  Note that 19\% and 69\% of the local DOGs and the comparison galaxies 
  do not have the MPA masses
  because they are not in a spectroscopic sample of SDSS DR7.}
  ratio $M_{\rm dust}/M_{\rm star}$ and 
  the mass fraction of warm dust component $M_{\rm warm}/M_{\rm dust}$ 
  are weakly correlated with infrared luminosity.

In the top and middle panels of Figure \ref{fig-tdust},
  both local DOGs and comparison galaxies show that 
  $T_{\rm cold}$ is in a narrow range 20--40 K, 
  but $T_{\rm warm}$ is a wide range 46--122 K. 
The K-S and A-D k-sample tests cannot reject the null hypothesis that 
  the dust temperature distributions between the two samples
  are drawn from the same parent population.
The bottom panels display the flux density ratios of 
  \iras 25 to 60 $\mu$m, $S_{25}/S_{60}$, 
  which is a model-independent warm dust temperature indicator (see \citealt{mag13}). 
Two samples show no significantly different distributions.
The $S_{25}/S_{60}$ are closely related to the AGN contribution and the warm dust temperature 
  in the sense that AGN-dominated galaxies usually have larger $S_{25}/S_{60}$ 
  (see green and blue symbols; \citealt{deg85,vei09,lee12}).
For comparison, we also plot the {\it IRAS}-detected SDSS galaxies at $z>0.01$ 
  regardless of submillimeter detection (gray dots and contours; \citealt{hwa10a}). 
The $S_{25}/S_{60}$ of both local DOGs and comparison galaxies 
  are slightly smaller than those of {\it IRAS}-detected SDSS galaxies 
  probably because the formers are relatively bright at long wavelengths 
  (i.e., they are submillimeter-detected; see \citealt{mag10}).

In the top panels of Figure \ref{fig-mdust},
  the dust masses of local DOGs seem to be on average larger than 
  those of comparison galaxies, 
However when we plot the dust masses normalized by stellar masses 
  ($M_{\rm dust}/M_{\rm star}$) in the middle panels, 
  we do not see such a difference.
The statistical tests also confirm that 
  overall dust masses of local DOGs are indistinguishable 
  from those of other infrared luminous galaxies.
In the bottom panels,
  there is no significant difference in the mass fraction of warm dust component 
  ($M_{\rm warm}/M_{\rm dust}$) between the two samples 
  ($p_{\rm KS}$ = 0.18 and $p_{\rm AD}$ = 0.19).
However, if we exclude the galaxies with a relatively large AGN contribution 
  (i.e., $f_{\rm AGN}>10\%$,
  the warm dust parameters of these galaxies could be contaminated by AGNs), 
  the difference between the two samples is significant 
  ($p_{\rm KS}$ = 0.02 and $p_{\rm AD}$ = 0.03).
Note that when the AGN-dominated galaxies are excluded, 
  the $p$-values change dramatically only for the warm dust mass fraction.

\section{DISCUSSION AND SUMMARY}\label{discuss}

We conducted CSO observations of 350 $\mu$m dust continuum emission for 17 local DOGs. 
By excluding three non-detected DOGs and 
  by including two DOGs with submillimeter data from the literature, 
  we derived their physical parameters of dust content 
  from the SED fits with a two-component dust model. 
Comparisons between the local DOGs and other submillimeter-detected infrared luminous galaxies
  show no significant difference in cold and warm dust temperatures and in total dust mass.

The two-component dust model in this study is 
  composed of two emissivity-modified blackbody functions, 
  representing the warm and cold dust components.
Although there are more sophisticated models that account for 
  various dust parameters (e.g., \citealt{dra07,dac08}), 
  this two-component model is good enough to 
  fit to the photometric data points at $\lambda > 20$ $\mu$m; 
  this does not include the mid-infrared part 
  where the contributions of polycyclic aromatic hydrocarbon (PAH) emission features and 
  stochastic heating processes are important (see \citealt{wal11} for a review). 
Actually, HAG13 show that dust masses derived from the two-component model 
  agree well with those from \citet{dra07} model (but see also \citealt{ber16}). 
\citet{kir15} also conclude that 
  the optically thin two-component modified blackbody model adopted in this study 
  is a reasonable choice for determining dust parameters
  compared to other methods including optically thick dust, fixed dust temperatures,
  and single-/three-component model.

According to the definition of \citet{dey08},
  the 16 local DOGs in this study can be classified into 4 power-law and 12 bump types.
Remarkably, all power-law DOGs have $f_{\rm AGN} > 20\%$ and 
  all bump DOGs have $f_{\rm AGN} \leq 20\%$. 
In the comparison sample of 35 galaxies with infrared luminosities similar to DOGs, 
  there is only one galaxy with $f_{\rm AGN} > 20\%$.
When we consider the AGN classification result of HG13 based on 
  optical emission line ratio diagrams \citep{bal81,vei87,kew06},
  the number fractions of AGNs among local DOGs and other galaxies are 
  51.1$\pm$7.3\% and 21.3$\pm$3.4\%, respectively. 
These confirm an important role of AGNs in local DOGs as an energy source,
  especially for power-law DOGs.

The SEDs of warm dust and AGN components 
  overlap significantly in the mid-infrared wavelengths. 
Because we consider each component separately, 
  dust parameters related to warm component could be 
  strongly affected by the AGN contamination. 
For example, the galaxies with $f_{\rm AGN} > 10\%$ are outliers 
  in the plot of $S_{25}/S_{60}$ vs. $L_{\rm IR}$ (bottom left panel of Figure \ref{fig-tdust}), 
  suggesting their warm dust parameters may not be reliable. 
There are several SED fitting codes that account for dust and AGN emission simultaneously 
  (e.g., \citealt{ber13,cie15}),
  but these codes are not useful to this study 
  because we would like to split the dust emission heated by star formation 
  into warm and cold components. 
The AGN contamination is significant only in the mid-infrared wavelengths 
  (e.g., \citealt{ass10,lee13}), 
  but not in the far-infrared/submillimeter wavelengths (e.g., \citealt{hat10,lee11,hwa11,kir12}). 
Therefore, the cold dust related parameters are not affected much even in AGN-dominated galaxies. 
The total dust mass is also unaffected because of very small fraction of warm dust mass 
  ($M_{\rm warm}/M_{\rm dust} \lesssim 1\%$).

Figures \ref{fig-tdust} and \ref{fig-mdust} show that 
  the AGN-dominated DOGs (with $f_{\rm AGN} > 10\%$) differ 
  from the starburst-dominated DOGs (with $f_{\rm AGN} \leq 10\%$) and 
  from comparison galaxies even in the distributions of 
  $T_{\rm cold}$, $M_{\rm dust}$, and $M_{\rm dust}/M_{\rm star}$ 
  where we expect little AGN effects. 
Statistical tests for the three samples suggest that the difference 
  in $T_{\rm cold}$ is insignificant, 
  but is significant in $M_{\rm dust}$ and $M_{\rm dust}/M_{\rm star}$
  (i.e., the dust masses of AGN-dominated DOGs are small.). 
However, the limited numbers of AGN- and starburst-dominated DOGs in this study 
  do not allow us to draw strong conclusions on the difference and its cause.

The bottom panels of Figure \ref{fig-mdust} show that 
  the amount of warm dust in starburst-dominated local DOGs is three times larger 
  than that in comparison galaxies; 
  when we exclude the galaxies with $f_{\rm AGN} > 10\%$, 
  the median values of $M_{\rm warm}/M_{\rm dust}$ are 0.62\% and 0.21\% 
  for local DOGs and comparison galaxies, respectively. 
Because the DOGs tend to be mid-infrared bright by definition, 
  the large fraction of warm dust component in DOGs could be expected. 
However, the main reason for the extreme near-UV/mid-infrared flux density ratios 
  of local DOGs is the abnormal faintness in the near-UV 
  rather than the mid-infrared brightness (see \citealt{pen12}; HG13).
It is thus not easy to understand why local DOGs have large warm dust fractions.
Moreover, the warm dust fraction range of comparison galaxies is very broad (2.6 dex). 
Because we cannot rule out the possibility that the derived warm dust parameters 
  are contaminated by the presence of (hidden) AGNs even in starburst-dominated DOGs, 
  this finding needs to be confirmed with a method free from AGN contamination. 
Then, with a larger sample of DOGs, it is also necessary to examine what physical properties 
  (e.g., stellar population, interstellar medium condition, extinction curve) are connected to this result.

On the other hand, the total dust masses of local DOGs are similar to 
  those of other submillimeter-detected infrared luminous galaxies. 
Interestingly, the (cold) dust masses of hot DOGs at $z\gtrsim2$ 
  (i.e., extreme AGN-dominated DOGs) appear comparable to 
  those of radio-detected quasars \citep{wu14}. 
Considering the large uncertainty in their dust mass estimates, 
  Wu et al. suggest that the dust masses of hot DOGs could be also similar to
  those of submillimeter galaxies (SMGs). 
The similar dust masses between DOGs and other galaxies 
  at both low and high redshifts indicate that 
  what makes DOGs special among dusty galaxies is 
  not an overall amount of dust content.
The other possible explanation for the extreme dust obscuration in local DOGs
  is that the dust distribution is spatially compact 
  such as centrally concentrated and/or clumpy 
  (strongly aligned with massive star-forming regions).
To investigate the detailed dust distribution in local DOGs, 
  we plan to observe the local DOGs with Atacama Large Millimeter Array (ALMA)
  and to look into the $Spitzer$ archival images of very nearby DOGs. 
We are also studying the gas content of local DOGs 
  from the James Clerk Maxwell Telescope (JCMT) observations, 
  which can provide valuable information on their dust-to-gas ratios, star formation efficiency, 
  and deviation from star-forming main sequence (see \citealt{dad10,elb11}).
  

\acknowledgments

We thank the anonymous referee for his/her useful comments that improved the manuscript.
We also thank Margaret Geller and Sean Andrews for helpful comments
in early stages of this work.
J.C.L. is a member of Dedicated Researchers for Extragalactic AstronoMy
(DREAM) in Korea Astronomy and Space Science Institute (KASI).
G.H.L. acknowledges the support by the National Research Foundation of Korea (NRF) Grant 
funded by the Korean Government 
(NRF-2012-Fostering Core Leaders of the Future Basic Science Program).
This material is based upon work at the Caltech Submillimeter Observatory,
which is operated by the California Institute of Technology.

\bibliography{} 


\end{document}

%% file: table1.tex
\begin{deluxetable*}{cccccllll}
\tabletypesize{\footnotesize}
\tablewidth{150mm}
\tablecaption{CSO Observation Log
\label{tab-samp}}
\tablehead{
ID                  & SDSS ObjID (DR9) & R.A.$_{2000}$ & Decl.$_{2000}$ & z & UT Date & $\tau_{225}$ & Int.  & $S_{350}$ \\
                    &                  &               &                &   &         &              & (min) & (mJy)       }
\startdata
 LDOG$-$07 &    1237674462024106294 & 09:04:01.02 & $+$01:27:29.12 & 0.0534 & 2014 Mar 19 & 0.088 & 28.4 & 475.9$\pm$86.0  \\
 LDOG$-$09 &    1237663530802937978 & 09:38:19.17 & $+$64:37:21.26 & 0.0710 & 2014 Mar 19 & 0.084 & 48.0 & 650.0$\pm$216.5  \\
 LDOG$-$14 &    1237654605860110514 & 10:17:31.29 & $+$04:36:19.04 & 0.0572 & 2014 Apr 10 & 0.063 & 38.2 & 180.0$\pm$31.9  \\
 LDOG$-$22 &    1237671141477777656 & 11:29:56.35 & $-$06:24:20.48 & 0.0523 & 2014 Apr 7  & 0.054 & 19.1 & 538.2$\pm$146.7  \\
 LDOG$-$23 &    1237657611801657347 & 11:35:49.09 & $+$56:57:08.27 & 0.0514 & 2014 Mar 19 & 0.065 & 76.2 & 257.3$\pm$85.7  \\
 LDOG$-$26 &    1237667209992732748 & 12:21:34.35 & $+$28:49:00.12 & 0.0613 & 2014 Apr 7  & 0.102 & 57.2 & 312.8$\pm$83.2  \\
 LDOG$-$27 &    1237667736660017246 & 12:56:25.47 & $+$23:20:55.05 & 0.0742 & 2014 Apr 8  & 0.056 & 37.9 & 327.5$\pm$48.1  \\
 LDOG$-$28 &    1237665129084092587 & 12:56:42.72 & $+$35:07:29.92 & 0.0547 & 2014 Apr 8  & 0.043 & 38.0 & 210.4$\pm$32.9  \\
 LDOG$-$33 &    1237665430241149030 & 13:41:02.95 & $+$29:36:42.86 & 0.0773 & 2014 Apr 10 & 0.061 & 47.3 & 220.4$\pm$33.7  \\
 LDOG$-$39 &    1237648705135051235 & 15:26:37.67 & $+$00:35:33.50 & 0.0507 & 2014 Apr 10 & 0.057 & 76.4 & 114.1$\pm$22.3  \\
 LDOG$-$41 &    1237662663216070833 & 15:51:53.04 & $+$27:14:33.65 & 0.0589 & 2014 Apr 10 & 0.070 & 16.4 & 345.9$\pm$43.3  \\
 LDOG$-$44 &    1237661387621073252 & 16:53:37.16 & $+$30:26:09.76 & 0.0732 & 2014 Apr 8  & 0.058 & 37.9 & 341.1$\pm$31.2  \\
 LDOG$-$45 &    1237668681527132368 & 17:03:30.38 & $+$45:40:47.15 & 0.0604 & 2014 Apr 10 & 0.049 & 48.0 & 157.0$\pm$28.1  \\
 LDOG$-$46 &    1237656530531254308 & 17:38:01.52 & $+$56:13:25.81 & 0.0652 & 2014 Apr 10 & 0.095 & 47.6 & 131.4$\pm$27.0  \\ 
\\[-7pt]\cline{1-9}\\[-7pt]
 LDOG$-$13 &    1237661383848951984 & 10:11:01.09 & $+$38:15:19.74 & 0.0527 & 2014 Apr 8  & 0.066 & 76.1 & $<$91.3  \\
 LDOG$-$16 &    1237648722831868077 & 10:33:33.15 & $+$01:06:35.15 & 0.0657 & 2014 Apr 10 & 0.060 & 57.3 & $<$203.3  \\
 LDOG$-$19 &    1237651067886502124 & 11:02:13.01 & $+$64:59:24.86 & 0.0776 & 2014 Mar 19 & 0.064 & 84.4 & $<$200.2
\enddata
\end{deluxetable*}

%% file: table2.tex
\begin{deluxetable*}{crrrrrrr}
\tabletypesize{\normalsize}
\tablewidth{120mm}
\tablecaption{SED Fit Results\label{tab-sed}}
\tablehead{
ID & $L_{\rm IR}$                & $f_{\rm AGN}$        &       & $T_{\rm warm}$     & $T_{\rm cold}$      & $M_{\rm dust}$            &  $M_{\rm warm}/M_{\rm dust}$  \\
   & ($\times10^{11} L_\odot$)   & (\%)                 &       & (K)                & (K)                 & ($\times10^{8} M_\odot$)  &  ($\times 10^{-3}$)              
}
\startdata
LDOG$-$07 & 4.49$^{+0.02}_{-0.02}$ &  0.0$^{+0.0}_{-0.0}$ & $~~~$ & 62.9$^{+1.9}_{-1.2}$ &  26.6$^{+0.2}_{-0.2}$ & 2.46$^{+0.04}_{-0.04}$ &  3.5$^{+0.8}_{-0.8}$ \\
LDOG$-$09 & 5.43$^{+0.07}_{-0.07}$ &  0.0$^{+0.0}_{-0.0}$ & $~~~$ & 80.1$^{+21.}_{-9.2}$ &  30.9$^{+1.3}_{-2.1}$ & 1.45$^{+9.99}_{-0.33}$ &  0.7$^{+1.7}_{-0.4}$ \\
LDOG$-$14 & 1.60$^{+0.04}_{-0.04}$ &  8.2$^{+0.7}_{-0.7}$ & $~~~$ & 60.7$^{+1.5}_{-1.1}$ &  23.1$^{+12.}_{-0.9}$ & 1.11$^{+0.31}_{-0.48}$ &  6.2$^{+2.3}_{-2.8}$ \\
LDOG$-$22 & 4.63$^{+0.05}_{-0.05}$ &  3.1$^{+0.3}_{-0.4}$ & $~~~$ & 58.9$^{+15.}_{-1.0}$ &  24.7$^{+3.5}_{-1.1}$ & 2.41$^{+0.87}_{-0.89}$ &  7.9$^{+2.0}_{-2.6}$ \\
LDOG$-$23 & 3.48$^{+0.11}_{-0.11}$ & 23.9$^{+1.9}_{-2.0}$ & $~~~$ &108.1$^{+15.}_{-8.6}$ &  35.0$^{+1.6}_{-1.5}$ & 0.27$^{+0.08}_{-0.06}$ &  0.8$^{+0.8}_{-0.4}$ \\
LDOG$-$26 & 2.74$^{+0.31}_{-0.58}$ &  0.0$^{+0.0}_{-0.0}$ & $~~~$ & 59.1$^{+4.8}_{-1.4}$ &  25.7$^{+9.5}_{-1.0}$ & 1.44$^{+0.27}_{-0.52}$ & 10.5$^{+3.9}_{-5.0}$ \\  
LDOG$-$27 & 3.94$^{+0.05}_{-0.05}$ &  0.0$^{+0.0}_{-0.0}$ & $~~~$ & 56.7$^{+28.}_{-0.7}$ &  22.2$^{+3.0}_{-0.9}$ & 3.42$^{+0.78}_{-0.82}$ &  9.6$^{+2.9}_{-2.1}$ \\
LDOG$-$28 & 2.39$^{+0.05}_{-0.03}$ &  0.0$^{+0.3}_{-0.0}$ & $~~~$ & 58.0$^{+99.}_{-0.6}$ &  22.7$^{+17.}_{-1.4}$ & 1.10$^{+0.35}_{-0.66}$ & 25.2$^{+7.9}_{-14.}$ \\
LDOG$-$33 & 3.76$^{+0.05}_{-0.04}$ &  0.0$^{+0.0}_{-0.0}$ & $~~~$ & 61.0$^{+2.1}_{-0.9}$ &  24.1$^{+8.6}_{-1.1}$ & 2.03$^{+0.51}_{-0.67}$ & 11.3$^{+3.6}_{-3.7}$ \\
LDOG$-$39 & 2.46$^{+0.07}_{-0.07}$ & 17.0$^{+1.3}_{-1.4}$ & $~~~$ &122.2$^{+39.}_{-17.}$ &  32.4$^{+1.4}_{-1.3}$ & 0.25$^{+0.07}_{-0.06}$ &  0.3$^{+0.7}_{-0.2}$ \\
LDOG$-$41 & 2.76$^{+0.08}_{-0.09}$ &  2.1$^{+0.6}_{-0.6}$ & $~~~$ & 58.5$^{+0.8}_{-0.6}$ &  22.8$^{+0.9}_{-0.8}$ & 2.24$^{+0.46}_{-0.41}$ &  6.1$^{+1.5}_{-0.9}$ \\
LDOG$-$44 & 4.71$^{+0.06}_{-0.06}$ &  0.0$^{+0.5}_{-0.0}$ & $~~~$ & 66.3$^{+75.}_{-2.5}$ &  26.0$^{+2.4}_{-0.9}$ & 2.43$^{+0.44}_{-0.70}$ &  3.3$^{+1.5}_{-2.0}$ \\
LDOG$-$45 & 3.13$^{+0.07}_{-0.07}$ & 44.4$^{+1.7}_{-1.7}$ & $~~~$ & 74.3$^{+26.}_{-1.3}$ &  23.8$^{+11.}_{-1.2}$ & 0.91$^{+0.31}_{-0.50}$ &  5.1$^{+8.0}_{-1.7}$ \\
LDOG$-$46 & 3.51$^{+0.09}_{-0.09}$ & 39.8$^{+2.1}_{-2.0}$ & $~~~$ & 98.4$^{+5.8}_{-8.6}$ &  32.1$^{+1.5}_{-3.0}$ & 0.40$^{+0.12}_{-0.15}$ &  1.2$^{+9.9}_{-0.4}$
\enddata
\end{deluxetable*}